\documentclass[12pt]{iopart}

\begin{document}

\title[Integrable $XY$ central spin models ]{Supersymmetry and integrability for a class of $XY$ central spin models}

\author{W J P van Tonder and J Links$^*$ }

\address{School of Mathematics and Physics, The University of Queensland, 4072, Australia}
\ead{$^*$jrl@maths.uq.edu.au}
\vspace{10pt}
\begin{indented}
\item[]%March 2023
\end{indented}

\begin{abstract}
Several studies have exploited the integrable structure of central spin models to deepen understanding of these fundamental systems. In recent years, an underlying supersymmetry for systems with $XX$ interactions has been uncovered. Here we report that a class of central spin models with $XY$ interactions is also supersymmetric and integrable. The associated Bethe Ansatz solution is presented for the case where all particles are spin-1/2. 
\end{abstract}

%\maketitle

\section{Introduction}

Central spin models increasingly draw attention for their potential applications in developing quantum technologies.
This interest is further driven by studies of their integrability, offering high-fidelity
control of mesoscopic quantum systems where the exponentially increasing size of the Hilbert space
makes such control challenging \cite{dldwlrd19}. The central spin allows the dynamics of the spin bath to be
monitored and for feedback to be used to steer the dynamics of the bath in a desired manner. A
physical realisation is accommodated by nitrogen vacancy centres in diamonds where one carbon atom is replaced
by a nitrogen atom and an adjacent carbon atom is absent. This acts like a spin-1/2 particle interacting
with a bath of nuclear spins from neighbouring carbon atoms \cite{yjgmgcl12}. An application exploiting
this high degree of control and robustness of the system is memory in quantum computers. The special
eigenstates of some of the central spin models provide a means of storing the qubit state of the central
spin among the bath spins and recovering it later \cite{til03,vcc20,vcppc20}. Other potential applications include quantum sensing
and metrology \cite{slcwpl14}, and utilising a central spin system with several central spins as a model of a quantum battery \cite{lsswy21}. In this latter setting the central spins serve as battery cells, that are charged by the bath spins.

As mentioned, much of the theoretical interest in central spin models stems from the existence of exact solutions, see e.g. \cite{bs07,fs13,g76,hclg22,jfnm09,ng18,w18,wh23}, and there are ongoing efforts to extend the body of known results. In recent times it was shown that a central spin-1/2 particle interacting with arbitrary bath spins is integrable for $XX$ interactions \cite{vcc20}. The result holds when the central spin is subjected to a magnetic field perpendicular to the plane of interaction. A surprising feature of this analysis was the appearance of {\it supersymmetry}, reminiscent of that observed in a class of $XY$Z spin chains \cite{hf12}. A complementary study showed that integrability persists for $XX$ interactions when an arbitrarily oriented magnetic field is applied to the central spin \cite{fs13}. However, the methods of the latter study were valid only for a bath of spin-1/2 particles. Here, we unify these approaches and extend them to establish that integrability holds for a class of $XY$ interactions, with arbitrary magnetic field applied to a spin-1/2 central spin and with arbitrary bath spins. Similar to \cite{vcc20}, we also identify a supersymmetric structure underlying the Hamiltonian.

In Sect. 2 we give the Hamiltonian of the model and express it in terms of conjugate supercharges. This formulation shows that the square of the Hamiltonian is, up to a constant, a supersymmetric operator. In Sect. 3 we then provide an extensive set of operators that commute with the square of the Hamiltonian. Restricting to the case where the bath consists entirely of spin-1/2 particles, we derive a Bethe Ansatz solution for the energies of the Hamiltonian in Sect. 4. Concluding remarks are provided in Sect. 5. 

\section{The Hamiltonian and supersymmetry}

Consider a set of $L+1$ spin operators $\{S^x_j,\,S_j^y,\,S^z_j: j=0,\dots L\}$ satisfying the standard canonical commutation relations
\begin{eqnarray}
&[ S^\alpha_j,\,S^\beta_k]=i\delta_{jk} \sum_{\gamma\in\{x,y,z\}}\varepsilon^{\alpha\beta\gamma} S^\gamma_j ,
\label{cr}
\end{eqnarray}
where $\varepsilon^{\alpha\beta\gamma}$ is the Levi-Civita symbol. We identify the spin labelled by 0 as the {\it central spin} and define a central spin Hamiltonian $H$ with $XY$ interactions to take the form
\begin{eqnarray}
H=B^x S_0^x+ B^y S_0^y +B^z S_0^z+\sum_{j=1}^L ( X_{j} S_0^x S_j^x+ Y_{j} S_0^y S_j^y).
\label{ham}
\end{eqnarray}
We fix the central spin to be spin-1/2, while the $L$ remaining {\it bath spins} have arbitrary spin. Without loss of generality, we may assume that all coupling parameters $B^x,\,B^y,\,B^z_0,\,X_j,\,Y_j$ are non-negative since the commutation relations (\ref{cr}) are invariant under parity-time transformations of the form $i\mapsto -i$, $\alpha\mapsto -\alpha$ for $\alpha\in\{x,\,y,\,z\}$.

We introduce a set of distinct parameters $\{\beta\}\cup\{\epsilon_j:j=1,\dots,L\}$ such that $\epsilon_j-\beta>0$ for all $j=1,\dots,L$ and we set $f_j^\pm =\sqrt{\epsilon_j\pm\beta}$. As is usual, define 
\begin{eqnarray*}
S_0^\pm =S_0^x\pm i S_0^y.
\end{eqnarray*} 
Next we introduce the {\it supercharges}
\begin{eqnarray}
\mathcal{A}^\pm=S_0^\mp\left( (\gamma\pm i\lambda)I+\sum_{j=1}^L (f_j^-S_j^x\pm if_j^+S^y_j)\right) ,
\label{sc}
\end{eqnarray}
where $I$ denotes the identity operator. These operators satisfy $(\mathcal{A}^\pm)^2=0$ and 
$S_0^z \mathcal{A}^\pm = -\mathcal{A}^\pm S_0^z$. It is straightforward to check that
\begin{eqnarray*}
\mathcal{A}^++\mathcal{A}^-&=2\gamma S_0^x+2\lambda S_0^y+ 2\sum_{j=1}^L (f_j^-S_0^xS_j^x+ f_j^+ S_0^yS^y_j)\\
&= H - 2\mu S_0^z
\end{eqnarray*} 
with the identification 
\begin{eqnarray*}
B^x&=2\gamma, \qquad Y_j^2+X_j^2&=2\epsilon_j, \\
B^y&=2\lambda, \qquad Y_j^2-X_j^2&=2\beta, \\
B^z&=2\mu.
\end{eqnarray*}
Moreover
\begin{eqnarray}
H^2&= \mu^2I+ (\mathcal{A}^++\mathcal{A}^-)^2 \nonumber\\
&= \mu^2I+\mathcal{A}^+ \mathcal{A}^- +\mathcal{A}^-\mathcal{A}^+  \nonumber\\
&= (\gamma^2+\lambda^2+\mu^2)I+Q
\label{hsq}
\end{eqnarray}
where 
\begin{eqnarray*}
Q=-2S^z_0\sum_{j=1}^L f_j^+f_j^- S_j^z&+2\sum_{j=1}^L(\gamma f^-_j S_j^x+\lambda f^+_j S_j^y) \\
&+\sum_{j,k=1}^L(f_j^-f_k^-S_j^x S_k^x+ f_j^+f_k^+ S_j^y S_k^y).
\end{eqnarray*}
The set of operators $\{H^2-\mu^2I,\,{\mathcal A}^\pm\}$ provide a realisation of the $sl(1|1)$ superalgebra and gives an example of a supersymmetric quantum mechanical system. Standard arguments show that the spectrum of $H^2-\mu^2I$ is non-negative, and that the non-zero energy eigenstates appear as ``boson/fermion'' pairs related through the action of ${\mathcal A}^\pm$ \cite{cks95}.  The next step is to establish that there exists a set of $L$ mutually commuting  operators that commute with $Q$, leading to the claim that the system is also integrable. 

\section{Integrability}
To expose the integrability of the system we take the approach to identify a set of mutually commuting operators that generalise Gaudin operators \cite{g76}, for arbitrary spins. Define 
\begin{eqnarray} 
	Q^\pm_j  &= \pm S_j^z  + \frac{2\gamma}{f^+_j} S_j^x + \frac{2\lambda}{f^-_j} S_j^y 
	+\frac{f_j^-}{f_j^+} (S_j^x)^2+\frac{f_j^+}{f_j^-} (S_j^y)^2 \nonumber \\
	&\qquad + 2\sum_{k\neq j}^L \frac{1}{\epsilon_j - \epsilon_k}\left( f^+_j f^-_k S_j^x S_k^x + f^-_j f^+_k S_j^y S_k^y \right) \nonumber \\
	&\qquad  + 2 \sum_{k \neq j}^L \frac{f^+_k f^-_k}{\epsilon_j - \epsilon_k} \left( S_j^z S_k^z - \frac{1}{4}I \right).
	\label{co}
\end{eqnarray}
Using the commutation relations (\ref{cr}) and the identities
\begin{eqnarray*}
0&=\frac{f_i^\mp f_j^\pm}{(\epsilon_i-\epsilon_j)} \frac{  f_j^\mp  f_k^\pm}{(\epsilon_j-\epsilon_k)} 
- \frac{f_i^\mp f_k^\pm}{(\epsilon_i-\epsilon_k)} \frac{f_j^+f_j^-}{(\epsilon_j-\epsilon_k)} -
\frac{ f_i^\mp f_k^\pm}{(\epsilon_i-\epsilon_k)} \frac{f_j^+ f_j^-}{(\epsilon_i-\epsilon_j)} \\
&=\frac{f_j^+ f_j^-}{(\epsilon_i-\epsilon_j)} \frac{ f_j^\mp f_k^\pm }{(\epsilon_j-\epsilon_k)} 
- \frac{f_k^+f_k^-}{(\epsilon_i-\epsilon_k)} \frac{ f_j^\pm f_k^\mp}{(\epsilon_j-\epsilon_k)} -
\frac{f_i^\mp f_k^\pm}{(\epsilon_i-\epsilon_k)} \frac{f_i^\mp f_j^\pm}{(\epsilon_i-\epsilon_j)} 
\end{eqnarray*}
it can be shown by direct calculation that 
\begin{eqnarray*}
[Q^\pm_j,\,Q_k^\pm] =0, \qquad j,k\in\{1,\dots, L\},
\end{eqnarray*}
generalising the results of \cite{l17a,vcc20} and being a specific case of those in \cite{s23}. Following \cite{vcc20} further, next set
\begin{eqnarray} 
	Q_j  &= \frac{1}{2}(I-2S_0^z)Q_j^+ + \frac{1}{2}(I+2S_0^z)Q_j^- \\
	&= -2S_0^z S_j^z  + \frac{2\gamma}{f^+_j} S_j^x + \frac{2\lambda}{f^-_j} S_j^y 
	+\frac{f_j^-}{f_j^+} (S_j^x)^2+\frac{f_j^+}{f_j^-} (S_j^y)^2 \nonumber \\
	&\qquad + 2\sum_{k\neq j}^L \frac{1}{\epsilon_j - \epsilon_k}\left( f^+_j f^-_k S_j^x S_k^x + f^-_j f^+_k S_j^y S_k^y \right) \nonumber \\
	&\qquad  + 2 \sum_{k \neq j}^L \frac{f^+_k f^-_k}{\epsilon_j - \epsilon_k} \left( S_j^z S_k^z - \frac{1}{4}I \right).
\end{eqnarray}
We find that 
\begin{eqnarray*}
Q=\sum_{j=1}^L f_j^+f_j^- Q_j
\end{eqnarray*}
from which $[H^2,\,Q_j]=0$ follows. This establishes that $H$ is an abstract integrable quantum system in the sense that the number of conserved operators for $H^2$ grows linearly with the number of spins. In the next section we will restrict to a bath of only spin-1/2 particles to illustrate how a Bethe Ansatz solution is obtained for the spectrum of $H$. 

\section{Bethe Ansatz solution for a bath of spin-1/2 particles}
For the case when all bath particles are spin-1/2 we have 
\begin{eqnarray*}
\left(S^x_j\right)^2=\left(S^y_j\right)^2=\frac{1}{4}
\end{eqnarray*}
in which case it is convenient to define the modified conserved operators 
\begin{eqnarray} 
\tilde{Q}_j&=	Q_j-\frac{1}{4}\left(\frac{f_j^-}{f_j^+}+\frac{f_j^+}{f_j^-}\right)I, \qquad j=1,...,L .
\label{mco}
\end{eqnarray}
It is known that the operators (\ref{mco}) 
satisfy a set of quadratic identities. For each simultaneous eigenstate of the operators given by (\ref{mco}), let $\{\tilde{q}_j\}$ denote the set of corresponding eigenvalues. These eigenvalues necessarily satisfy analogous quadratic relations that read \cite{cddf19} 
\begin{eqnarray} 
    \tilde{q}_j^2 
= \frac{1}{4} &+ \frac{\gamma^2}{(f^+_j)^2} + \frac{\lambda^2}{(f^-_j)^2} -  \sum_{k\neq j}^L f^+_k f^-_k \left( \frac{\tilde{q}_j - \tilde{q}_k}{\epsilon_j - \epsilon_k} \right) 
\nonumber \\
&+ \frac{1}{4}  \sum_{j\neq i}^L \left( \frac{f^+_j f^-_k - f^-_j f^+_k}{\epsilon_j-\epsilon_k} \right)^2    .
\label{quad}
\end{eqnarray}

To extract Bethe Ansatz solutions from the above quadratic relations we will adapt methods developed in \cite{l17b,l17c}. 
Set 
\begin{eqnarray*}
\alpha_\pm=\frac{1}{2}(L+1\pm 1).
\end{eqnarray*}
Also assume the eigenvalues of the conserved operators to have the form 
\begin{eqnarray}
\tilde{q}_j=\frac{\alpha_\pm \epsilon_j}{f_j^+f_j^-}
-\sum_{m=1}^L \frac{f_j^+f_j^-}{\epsilon_j-v_m}+\frac{1}{2}\sum_{k\neq j}^L\frac{f_j^+f_j^- - f_k^+f_k^-}{\epsilon_j-\epsilon_k}
\label{param}
\end{eqnarray}
so that the $\{q_j:j=1,\dots,L\}$ are parametrised in terms of variables $\{v_j:j=1,\dots,L\}$. Note that this is always possible to achieve. Setting
\begin{eqnarray*}
Q(u)= \prod_{j=1}^L(u-v_j)=u^L+\sum_{j=0}^{L-1} a_j u^j,
\end{eqnarray*}
such that 
\begin{eqnarray*}
\frac{Q'(u)}{Q(u)}=\sum_{j=1}^L \frac{1}{u-v_j},
\end{eqnarray*}
then (\ref{param}) provides a system of $L$ homogeneous linear equations for the $L+1$ coefficients $\{a_j\}$, and this system admits a non-trivial solution\footnote{If the solution gives $Q(u)$ as a polynomial of degree $M<L$ such that $a_L=0$, this is to be interpreted as $L-M$ of the roots being {\it infinite-valued}. See \cite{lmm15} for an example of this feature. }.

The form (\ref{param}) has been chosen in such a way that the relations (\ref{quad}), which are expressed in terms of irrational algebraic functions of the $\{\epsilon_j\}$, can be transformed into rational functions. Inserting (\ref{param}) into (\ref{quad}) leads to 
\begin{eqnarray*}
S(\epsilon_j)=0, \qquad j=1,\dots,L \label{zeroes}
\end{eqnarray*}
where
\begin{eqnarray*}
S(u)=&\alpha_\pm^2   \beta^2 -\gamma^2(u-\beta)-\lambda^2(u+\beta) \\
&+(u^2-\beta^2) \left((2L-2\alpha_\pm)\sum_{m=1}^L\frac{v_m}{u-v_m}+\sum_{j,m=1}^L\frac{v_m^2-\beta^2}{(u-v_m)(\epsilon_j-v_m)} \right) \\
&+2(u^2-\beta^2)\sum_{m=1}^L\sum_{n\neq m}^L    \frac{v_mv_n-\beta^2}{(u-v_m)(v_m-v_n)}.
\end{eqnarray*}
By observing that $Q(u)S(u)$ is a polynomial of degree $L+1$, and also using 
\begin{eqnarray*}
S(\beta)&=\alpha^2_\pm \beta^2-2\lambda^2\beta  , \\
S(-\beta)&=\alpha^2_\pm \beta^2+2\gamma^2\beta ,
\end{eqnarray*}
it follows from (\ref{zeroes}) that
\begin{eqnarray*}
S(u)=f(u)\frac{P(u)}{Q(u)}
\end{eqnarray*}
where 
\begin{eqnarray*}
P(u)=\prod_{j=1}^L(u-\epsilon_j)
\end{eqnarray*}
and 
\begin{eqnarray*}
f(u)=\frac{1}{2}\left( (\alpha_\pm^2\beta- 2\lambda^2)\frac{Q(\beta)}{P(\beta)}(u+\beta) - (\alpha_\pm^2\beta+2\gamma^2)\frac{Q(-\beta)}{P(-\beta)}(u-\beta)  \right).
\end{eqnarray*}
Evaluating  
\begin{eqnarray*}
\lim_{u\rightarrow v_l} (u-v_l) f(u)\frac{P(u)}{Q(u)}= \lim_{u\rightarrow v_l} (u-v_l)S(u)
\end{eqnarray*}
leads to the Bethe Ansatz equations
\begin{eqnarray}
&\frac{1}{2}\prod_{n\neq l}^L\frac{1}{v_l-v_n}\prod_{m=1}^L(v_l-\epsilon_m)\left(\frac{\alpha_\pm^2\beta-2\lambda^2}{v_l-\beta}\frac{Q(\beta)}{P(\beta)}-\frac{\alpha^2_\pm \beta+2\gamma^2}{v_l+\beta}
\frac{Q(-\beta)}{P(-\beta)}  \right) \nonumber \\
%&\qquad=(2L-2\alpha_\pm )v_l+\sum_{j=1}^L
%\frac{v_l^2-\beta^2}{\epsilon_j-v_l}+2\sum_{m\neq l}^L \frac{v_lv_m-\beta^2}{v_l-v_m} \nonumber \\
%&\qquad=(2L-2\alpha_\pm )v_l+\sum_{j=1}^L
%\frac{v_l^2-\epsilon_j^2+\epsilon_j^2-\beta^2}{\epsilon_j-v_l}+2\sum_{m\neq l}^L \frac{v_lv_m-\beta^2}{v_l-v_m} \\
%&\qquad=(2L-2\alpha_\pm )v_l-\sum_{j=1}^L
%(\epsilon_j+v_l)+\sum_{j=1}^L
%\frac{\epsilon_j^2-\beta^2}{\epsilon_j-v_l}+2\sum_{m\neq l}^L \frac{v_lv_m-\beta^2}{v_l-v_m} \\
&\qquad=(L-2\alpha_\pm )v_l-\sum_{j=1}^L
\epsilon_j+\sum_{j=1}^L
\frac{\epsilon_j^2-\beta^2}{\epsilon_j-v_l}+2\sum_{m\neq l}^L \frac{v_lv_m-\beta^2}{v_l-v_m} .
\label{bae}
\end{eqnarray}

Since $P(u)-Q(u)$ has degree less than $L$, Lagrange basis polynomials may be used to give 
\begin{eqnarray*}
P(u)-Q(u) &=\sum_{j=1}^L (P(v_j)-Q(v_j))\prod^L_{k\neq j}\frac{u-v_k}{v_j-v_k} \\
&=\sum_{j=1}^L P(v_j)\prod^L_{k\neq j}\frac{u-v_k}{v_j-v_k} .
\end{eqnarray*}
Then taking the sum over $l$ in (\ref{bae}) leads to 
\begin{eqnarray}
&\frac{1}{2}\left( (2\lambda^2-\alpha_\pm^2\beta) \left(1-\frac{Q(\beta)}{P(\beta)}\right)  +
   (2\gamma^2+\alpha_\pm^2\beta) \left(1-\frac{Q(-\beta)}{P(-\beta)}\right)  \right) \nonumber\\
&\qquad=\sum_{l=1}^L(L-2\alpha_\pm )v_l-L\sum_{j=1}^L
\epsilon_j+\sum_{l=1}^L\sum_{j=1}^L
\frac{\epsilon_j^2-\beta^2}{\epsilon_j-v_l} .
\label{insert}
\end{eqnarray}
The eigenvalue ${\mathcal Q}$ of $Q$ reads
\begin{eqnarray*}
{\mathcal Q}&=\sum_{j=1}^L f_j^+f_j^- \left(\tilde{q}_j +\frac{f_j^-}{4f_j^+}
 +\frac{f_j^+}{4f_j^-} \right)  \\ 
%&=\sum_{j=1}^L \left( \left(\alpha_\pm+\frac{1}{2}\right) \epsilon_j
%-\sum_{m=1}^L \frac{\epsilon_j^2-\beta^2}{\epsilon_j-v_m}+\frac{1}{2}\sum_{k\neq j}^L\frac{\epsilon_j^2-\beta^2 %- f_j^+f_j^-f_k^+f_k^-}{\epsilon_j-\epsilon_k} \right) \\
&= \frac{1}{2}(2\alpha_\pm +L) \sum_{j=1}^L \epsilon_j
-\sum_{j=1}^L \sum_{m=1}^L \frac{\epsilon_j^2-\beta^2}{\epsilon_j-v_m}.
\end{eqnarray*} 
Using (\ref{insert}) we obtain the squares of the energies, given by 
\begin{eqnarray}
E^2&= \gamma^2&+\lambda^2+\mu^2+{\mathcal Q} \nonumber\\
&=\mu^2 &+\sum_{l=1}^L(L-2\alpha_\pm)v_l  +\frac{1}{2}(2\alpha_\pm -L)  \sum_{j=1}^L\epsilon_j \nonumber \\
 &&+\frac{1}{2}\left( (2\lambda^2-\alpha_\pm^2\beta) \frac{Q(\beta)}{P(\beta)}  +
   (2\gamma^2+\alpha_\pm^2\beta) \frac{Q(-\beta)}{P(-\beta)}  \right) .
\label{nrg}
\end{eqnarray}
It is important to highlight that the Bethe Ansatz solution given by (\ref{bae},\ref{nrg}), for generic values of the coupling parameters, is complete and accounts for all energies of the Hamiltonian. The line of reasoning follows the same arguments as presented in \cite{l17b,l17c} and is based on the fact that the quadratic relations (\ref{quad}) obtained from operator identities are complete. But in addition it needs to be asserted that for each value of $E^2$ given by (\ref{nrg}), both the positive and negative values for $E$ appear in the spectrum  upon taking the square root. This follows from our earlier observation that the signs of the coupling parameters $\{B^x,\,B^y,\,B^z,\,X_j,\,Y_j\}$, appearing linearly in (\ref{ham}), can be changed by unitary transformations. 
%and can also be viewed through the prism of the existence of the supercharges.  

We remark finally that in the limit $\beta,\gamma,\lambda\rightarrow 0$ expressions (\ref{bae}) and (\ref{nrg}) coincide with those found in \cite{vcc20} for the $XX$ model, up to a change a variables. The solutions for the choice $\alpha_+$ correspond to the entangled {\it bright states}, while those  for the choice $\alpha_-$ correspond to the separable {\it dark states} as defined therein.  

\section{Discussion}

We have shown that the supersymmetry and integrability of the central spin-1/2 model with $XX$ interactions and arbitrary bath spins extends to a class of $XY$ interactions. We have explicitly identified the supercharges (\ref{sc}) and a set of $L$ mutually commuting conserved operators as given by (\ref{co}).  In the case where all bath particles are spin-1/2, we have used a set of known quadratic identities to derive a Bethe Ansatz solution.

For future work there are several avenues available.
A generalisation of the Bethe Ansatz results for arbitrary spins is in principle attainable following the tensor product methods constructed in \cite{l17a}. This is feasible because every higher spin with a finite-dimensional state space can be obtained through a tensor product of spin-1/2 spaces and an appropriate projection.

In the limit as $L\rightarrow\infty$ integral techniques can be employed to obtain an expression for the ground-state energy from the quadratic identities (\ref{quad}). This calculation was undertaken in \cite{sil20} for an analogous BCS pairing Hamiltonian. It is known that a  correspondence exists between BCS models and central spin models 
 \cite{yake05}, and this may be used to adapt and  translate the results of \cite{sil20} to the $XY$ central spin model (\ref{ham}) in the spin-1/2 bath case.

While the Bethe Ansatz approach described here does not yield expressions for the eigenstates,  these are in principle accessible by adapting the algebraic Bethe Ansatz approach developed in \cite{s22,s23}. This appears to be a highly technical challenge. But it would be very useful to gain a better physical understanding for the analogues of bright and dark states associated with the choice for $\alpha_\pm$, characterised in \cite{vcc20} for $XX$ interactions, within the $XY$ model.

Another path to follow is to generalise the studies  for higher-order central spins \cite{tlpcc23,wgl20}  with $XX$ interactions to the $XY$ setting. 

\ack{The authors acknowledge the traditional owners of
the land on which The University of Queensland at St. Lucia operates,
the Turrbal and Jagera people. This work was supported by the Australian Research Council through Discovery Project DP200101339.}

\end{document}